\documentclass{aa} 
 
\usepackage{psfig, times} 

\newcommand{\M}{{\sc 2mass}}

\begin{document} 
 
\title{The distance to the LMC cluster NGC~1866; clues from the
cluster Cepheid population}  
 
\author{ 
M.A.T. Groenewegen 
\inst{1}  
\and 
M. Salaris 
\inst{2}  
} 
  
\institute{ 
Instituut voor Sterrenkunde, PACS-ICC, Celestijnenlaan 200B,  
B--3001 Leuven, Belgium  
\and 
Astrophysics Research Institute, Liverpool John Moores University, 
Twelve Quays House, Birkenhead CH41 1LD, UK 
} 
 
\date{received: 2003,  accepted: 2003} 
 
\offprints{Martin Groenewegen (groen@ster.kuleuven.ac.be)} 
 
\authorrunning{Groenewegen \& Salaris} 
\titlerunning{The Cepheid distance to the LMC cluster NGC 1866}

\abstract{ 
Recent investigations aimed at estimating the distance to the young
LMC cluster NGC~1866 have made use of Red Clump stars in the
surrounding LMC field, together with empirical and theoretical
Main-Sequence fitting methods, and have found significantly different
distances for the field and the cluster, the latter being closer by in
distance modulus by $\Delta$(DM)$\sim$0.20 mag.  In this paper we
(re-)consider the Cepheid star population of NGC~1866, to try to shed
some light on this discrepancy. By combining various extensive
photometric datasets in $B$, $V, I$ and single-epoch \M\ $JHK$
photometry, $PL$ relationships for the cluster Cepheids are obtained.
A comparison between the field LMC and cluster $PL$ relationships for
the reddening free Wesenheit index gives a firm determination of the
distance between the cluster and the LMC main body (0.04 mag in
distance modulus, the cluster being more distant) which, coupled to a
model for the geometry of the LMC disk, provides $\Delta$(DM) ranging
between 0.0 and $-$0.11 mag. The simultaneous comparison of the $PL$
relationships in $B$, $V$ and $I$ for the cluster and LMC field gives
an estimate of the cluster reddening, which results to be $E(B-V)$ =
0.12 $\pm$ 0.02.  This determination is higher than the canonical
value of 0.06 mag used in all previous studies, but we show that it is
not in contradiction with a re-analysis of independent estimates. The
adoption of the LMC extinction law recently presented by Gordon et
al.~(2003) does not change these results. The cluster Main Sequence
fitting distance obtained with this new reddening is DM = 18.58 $\pm$
0.08, fully compatible with the Red Clump value of DM = 18.53 $\pm
0.07(random)^{+0.02}_{-0.05}(systematic)$ and the Cepheid constraint
on $\Delta$(DM).  Finally, we determined the distance to the cluster
by using a Cepheid Wesenheit $PL$ relationship with slope coming from
LMC observations, and absolute magnitude zero point calibrated on
$Hipparcos$ parallaxes of Galactic Cepheids, in the assumption that
the relationship is independent of metallicity; the resulting DM =
18.65 $\pm$ 0.10 is not an accurate estimate of the LMC distance
because of possible metallicity effects but, when compared to the
revised Main Sequence fitting value, it points out to a possibly weak
dependence of the Wesenheit $PL$ relationship on the Cepheid chemical
composition, at least in the period range between 2.5 and 3.5 days.
} 
 
 
\maketitle

\section{Introduction} 
 
Two recent papers considered the distance to the young LMC cluster
NGC~1866 based on empirical and theoretical Main-Sequence (MS) fitting
techniques, and on the Red Clump (RC) distance to the LMC field around
the cluster. Walker et al.~(2002) used a MS-fitting method employing
theoretical isochrones to derive a distance modulus (DM) to the
cluster DM = 18.35 $\pm$ 0.05, and a reddening of $E(B-V)$ =
0.060-0.064. Salaris et al. (2003a, hereafter S03) employed the same
cluster photometry to derive DM = 18.33 $\pm$ 0.08 (for an adopted
reddening of 0.064) using a completely empirical MS-fitting method,
based a large sample of local subdwarfs with accurate parallax and
[Fe/H] determination; however, when they applied the RC method
(following the procedure by Alves et al.~2000 and the population
corrections by Salaris \& Girardi~2002) for deriving the distance to
the surrounding LMC field, it resulted a distance modulus DM =
18.53 $\pm$ 0.07 (a similar value of 18.47 $\pm$ 0.05 is obtained by
Pietrzy\'nski \& Gieren~2002 using the RC in the $K$-band).  This DM
discrepancy $\Delta$(DM) = 0.20 $\pm$ 0.10 between RC and MS-fitting
distances reflects the more general dichotomy in the LMC distance
estimates found in the literature (see, e.g., Benedict et al.~2002 for
a recent summary of the LMC distance determinations).  S03 discussed
possible reasons for this occurrence, as an underestimated cluster
metallicity, a photometric zero point error, the possibility that the
cluster is located about 5 Kpc closer than the underlying field
population, but no definitive conclusion was reached.
 
NGC~1866 contains a sizable Cepheid population, and therefore an
independent distance estimate is potentially available.  Shapley \&
Nail~(1950) and Thackeray~(1951) independently discovered the first
Cepheids in NGC~1866, and first photometry and periods were published
by Arp \& Thackeray (1967; hereafter AT67); from Period-Luminosity
($PL$) and Period-Luminosity-Colour ($PLC$) relations with the then
available calibration they derived a DM to the cluster of 18.44 $\pm$
0.15 (for a reddening of $E(B-V)$ = 0.06).  After the publication by
Walker~(1987, hereafter Wa87) of the first CCD light curves for seven
cluster Cepheids, Storm et al.~(1988) report the discovery of 10 new
Cepheid candidates. Welch et al. (1991; hereafter We91) present new
CCD photometry and radial velocity measurements of these new and
previously known objects. For a reddening of $E(B-V)$ = 0.06, and an
assumed $PLC$-relation they derive DM = 18.57 $\pm$ 0.01 (internal
error only). These data were then used by C\^ote et al.~(1991) for a
Baade-Wesselink analysis to derive effective temperatures and radii,
and DM = 18.6 $\pm$ 0.3. Gieren, Richtler \& Hilker~(1994)
derived an improved DM = 18.47 $\pm$ 0.20 from the first $VRI$-based
Baade-Wesselink analysis of 3 cluster Cepheids.  Gieren et
al. (2000a; hereafter G2000) report extensive new $BVRI$ photometry
for seven Cepheids and provide improved periods, and at the same time
Gieren et al.~(2000b) discuss the application of the infrared surface
brightness Baade-Wesselink method on a Cepheid (HV 12198) in the
cluster. From this one star, they determine DM = 18.42 $\pm$ 0.10.
 
It is clear that today extensive photometry for many of the Cepheids
in NGC 1866 does exist. Single-epoch infra-red data are also
potentially available from the \M\ survey.  Current Cepheid-based
distance estimates date back to more than 10 years ago, i.e. in the
era before the Hipparcos based calibration of the ZP of the Galactic
$PL$-relation, and before the huge datasets of field Cepheids in the
Magellanic Clouds, discovered by the microlensing surveys.  The aim of
the present paper is therefore to combine all available data for the
Cepheids in NGC~1866 with our current knowledge of field LMC and
Galactic Cepheids, in order to shed some light on the dichotomy of
distances estimated from MS-fitting and the RC-method. In Sect.~2 we
discuss the available Cepheid photometric data; inferences from their
$PL$ relationships will be analysed in Sect.~3, and a discussion about
the implications for the cluster distance follows in Sect.~4.

\section{Photometry of NGC~1866 Cepheids} 
 
We considered the photometry ($V$, and when available $B$ and $I$)
from G2000, We91, Wa87 and AT67. As previously noted and discussed in
the relevant papers, there is for some stars a small difference
between the different sets of photometry (e.g., see the phase diagrams
by G2000).  Furthermore, the photometry in AT67 is made of
photographic magnitudes whereas data in the other papers are obtained
using CCDs. Lastly, G2000 and We91 quote (internal) errors while Wa87
and AT67 do not. On the other hand we want to use as much data as
possible with representative errors to make full use of all available
information. It should also pointed out that crowding is a
potential problem for the Cepheids closer to the centre of the cluster
(in particular the Cepheids with the $V$ prefix in their name) 
and new photometry under {\em excellent} seeing conditions would be valuable.
 
In order to do so we took a two-step approach. As a first step, we
fixed the periods to the most accurate known values (i.e. as quoted in
the paper based on the single dataset with most observations, hence
G2000 in most cases). For each dataset we then solved for the
amplitude and phase, using the numerical code ``Period98''
(Sperl~1998). We also considered lightcurves with a suitable number of
harmonics (typically 3 to 5), solving again for the amplitudes and
phases, the relevant output quantities being the mean magnitude and
the r.m.s. The difference between the mean magnitudes obtained with
different datasets gives an indication of the photometric offsets, and
the rms gives an indication of the error in an individual measurement.

Based on this exercise, an error of 0.008 mag was assigned to Wa87
data, and 0.05 to the $V$-band data in AT67. In addition, the
following offsets were added to the published photometries in order to
put them on the same system as G2000: +0.039 mag ($I$, Wa87,
HV~12197), +0.082 mag ($V$, We91, HV~12200), $-0.066$ mag ($I$, Wa87,
HV~12203), and to the photographic $V$-band in AT67 we added +0.095
mag for HV~12197, 12198, 12199, 12205, +0.212 mag for HV~12202 and
+0.079 mag for HV~12203.

In case of the $B$-band we added +0.04 mag (Wa87, HV~12197),
$-0.06$ mag (We91, HV~12197) +0.02 mag (Wa87, HV~12199), +0.02 mag
(We91, HV~12199), +0.15 mag (We91, HV~12200), $-0.02$ mag (Wa87,
HV~12202), +0.05 mag (We91, HV~12202), +0.06 mag (Wa87, HV~12203),
+0.03 mag (We91, HV~12203) and $-0.02$ mag (We91, V7).

HV~12204 was not considered as it is likely a non-member 
based on its radial velocity (Wa87). 
 
In a second step we combined all datasets (with offsets and proper
weighting applied) and performed a Fourier analysis solving for the
primary frequency, amplitudes and phases. The results are listed in
Tables~\ref{Tab-fourier-V}, \ref{Tab-fourier-B} and
\ref{Tab-fourier-I}. When fitting the $B$ and $I$-band lightcurve the
frequencies where fixed to that determined from the
$V$-lightcurves. Also listed is the rms in the final fit. This number
will be used in the next section to characterise the error in the mean
magnitude when using the $PL$-relationship.

In the case of the $B$-band, keeping the frequency also as a free
parameter resulted in a different frequency by (1-3) $10^{-6}$
cycles/day at most, and no significant change in the mean magnitude
and rms values. For the stars in common, the periods derived here and
those quoted in G2000 agree within their respective 2$\sigma$ error bars.

The quantities $R_{21} = A_2/A_1$ and $\phi_{21} = \phi_2 - 2
\phi_1$ (where $A_i$ and $\phi_i$ represent the amplitude and phase of
the (i-1)-harmonic in the Fourier expansion) can be used to
distinguish fundamental (FU) from first overtone (FO) pulsators
(e.g. Udalski et al. 1999b); when applying this technique we found
that Cepheid V8 is a first overtone pulsator (We87 already suggested
this purely on the basis of its short period). For V4 the situation is
ambiguous and it is kept as a FU pulsator. In case of V6, although
the amplitude ratio suggests it is a FU, the object is treated as a FO
pulsator for reasons given below (note that We87 also suggested it to
be a probable FO pulsator). When plotted on a $V$-band $PL$-diagram
(like Fig.~3 below), it would stand out as a clear outlier at its
observed period of 2.05 days, being about 0.4 mag brighter than the
mean relation at that period, a deviation by many sigmas. At the same
time, it falls almost exactly in the middle of datapoints of FO
pulsators in the LMC field at that period (Udalski et al. 1999a). In
addition, the number of field LMC Cepheids in the period range 2.03 $<
P <$ 2.07 $d$ is 0 FU and 15 FO, and in the extended range 2.00 $< P
<$ 2.10 $d$ is 4 FU and 33 FO. This statistical argument (in the
hypothesis that cluster and field Cepheids share the same properties),
and the fact that it is located on top of the $V$-band $PL$-relation
of LMC field FO Cepheids, make us believe that V6 is an overtone
pulsator. If this object is excluded from our analysis, the results we
present in the following are completely unaffected.  All other
Cepheids in our sample appear to be FU pulsators. 

For the FO variables V6 and V8 the observed period ($P_1$) has been transformed into the
corresponding fundamental one ($P_0$) according to (Feast \& Catchpole~1997):
\begin{equation} 
P_1/P_0 = 0.716 - 0.027 \log P_1 
\end{equation}

\begin{table} 
\caption[]{ Fourier decomposition of $V$-lightcurves} 
\label{Tab-fourier-V} 
\begin{tabular}{lllllll} \hline 
 Name & $<V>$  & Period ($d$) & rms   & N   & Ampl   & phase \\ 
\hline  
12197 & 16.102 & 3.14374(3)   & 0.028 & 100 & 0.226  & 0.399 \\ 
      &        &              &       &     & 0.079  & 0.208 \\ 
      &        &              &       &     & 0.0305 & 0.984 \\ 
      &        &              &       &     & 0.0137 & 0.736 \\ 
 
12198 & 15.976 & 3.52275(3) & 0.030 & 134 & 0.258  & 0.418 \\ 
      &        &            &       &     & 0.112  & 0.260 \\ 
      &        &            &       &     & 0.0553 & 0.113 \\ 
      &        &            &       &     & 0.0162 & 0.021 \\ 

12199 & 16.289 & 2.639166(2) & 0.059 & 113 & 0.270  & 0.963 \\ 
      &        &             &       &     & 0.127  & 0.330 \\ 
      &        &             &       &     & 0.0610 & 0.735 \\ 
      &        &             &       &     & 0.0349 & 0.087 \\ 
      &        &             &       &     & 0.0093 & 0.439 \\ 
 
12200 & 16.247 & 2.72499(3) & 0.042 & 63 & 0.325  & 0.767 \\ 
      &        &            &       &    & 0.153  & 0.927 \\ 
      &        &            &       &    & 0.0807 & 0.103 \\ 
      &        &            &       &    & 0.0599 & 0.316 \\ 
      &        &            &       &    & 0.0258 & 0.417 \\ 
 
12202 & 16.080 & 3.10118(1) & 0.037 & 122 & 0.211  & 0.691 \\ 
      &        &            &       &     & 0.0903 & 0.786 \\ 
      &        &            &       &     & 0.0349 & 0.914 \\ 
      &        &            &       &     & 0.0091 & 0.897 \\ 
 
12203 & 16.140 & 2.95414(2) & 0.028 & 115 & 0.252  & 0.962 \\ 
      &        &            &       &     & 0.106  & 0.340 \\ 
      &        &            &       &     & 0.0519 & 0.734 \\ 
      &        &            &       &     & 0.0232 & 0.119 \\ 
      &        &            &       &     & 0.0054 & 0.606 \\ 
 
12205 & 15.965 & 3.21048(2) & 0.090 & 32 & 0.330 & 0.728 \\ 
      &        &            &       &    & 0.143 & 0.862 \\ 
      &        &            &       &    & 0.122 & 0.117 \\ 
 
V4    & 16.068 & 3.31886(2) & 0.022 & 48 & 0.113 & 0.391 \\ 
      &        &            &       &    & 0.027 & 0.185 \\ 
 
V6    & 16.111 & 2.05442(2) & 0.030 & 36 & 0.080 & 0.207 \\ 
      &        &            &       &    & 0.049 & 0.049 \\ 
      &        &            &       &    & 0.049 & 0.203 \\ 
      &        &            &       &    & 0.019 & 0.132 \\ 
 
V7    & 15.973 & 3.38827(2) & 0.065 & 79 & 0.131  & 0.723 \\ 
      &        &            &       &    & 0.0364 & 0.827 \\ 
      &        &            &       &    & 0.0267 & 0.912 \\ 
 
V8    & 16.159 & 2.05249(2) & 0.036 & 46 & 0.128 & 0.033 \\ 
      &        &            &       &    & 0.047 & 0.624 \\ 
      &        &            &       &    & 0.021 & 0.898 \\ 
      &        &            &       &    & 0.016 & 0.584 \\ 
\hline 
\end{tabular} 

Listed are the identifier, mean magnitude, period in days with the
uncertainty in the last digit between parenthesis, the rms in the fit,
the number of datapoints, and then the amplitude and phase of the
Fourier components, one component in each line.  The solutions listed
for $B$ and $I$ in Tables~\ref{Tab-fourier-B} and \ref{Tab-fourier-I}
have been obtained with the frequency indicated here. Leaving the
frequency as a free parameter will lead to very small differences in
the derived quantities, that have been the basis for attributing an
error to the period.

\end{table}

\begin{table} 
\caption[]{ Fourier decomposition of $B$-lightcurves}
\label{Tab-fourier-B} 
\begin{tabular}{llllll} \hline 
 Name & $<B>$  & rms   & N  & Ampl   & phase \\ \hline  
12197 & 16.741 & 0.024 & 50 & 0.338  & 0.418 \\ 
      &        &       &    & 0.114  & 0.222 \\ 
      &        &       &    & 0.0545 & 0.041 \\ 
      &        &       &    & 0.0088 & 0.802 \\ 
 
12198 & 16.627 & 0.024 & 83 & 0.284  & 0.384 \\ 
      &        &       &    & 0.159  & 0.272 \\ 
      &        &       &    & 0.0790 & 0.138 \\ 
      &        &       &    & 0.0209 & 0.041 \\ 
 
12199 & 16.916 & 0.015 & 62 & 0.411  & 0.974 \\ 
      &        &       &    & 0.177  & 0.322 \\ 
      &        &       &    & 0.0833 & 0.723 \\ 
      &        &       &    & 0.0454 & 0.061 \\ 
      &        &       &    & 0.0153 & 0.4458 \\ 
 
12200 & 16.913 & 0.053 & 62 & 0.518  & 0.779 \\ 
      &        &       &    & 0.230  & 0.929 \\ 
      &        &       &    & 0.137  & 0.084 \\ 
      &        &       &    & 0.0594 & 0.299 \\ 
      &        &       &    & 0.0323 & 0.389 \\ 
 
12202 & 16.756 & 0.019 & 71 & 0.303  & 0.706 \\ 
      &        &       &    & 0.104  & 0.798 \\ 
      &        &       &    & 0.0442 & 0.912 \\ 
      &        &       &    & 0.0114 & 0.163 \\ 
 
12203 & 16.811 & 0.026 & 64 & 0.378  & 0.976 \\ 
      &        &       &    & 0.150  & 0.349 \\ 
      &        &       &    & 0.0919 & 0.723 \\ 
      &        &       &    & 0.0235 & 0.102 \\ 
      &        &       &    & 0.0195 & 0.608 \\ 

12205 & 16.572 & 0.013 & 11 & 0.557 & 0.748 \\ 
      &        &       &    & 0.222 & 0.836 \\ 
      &        &       &    & 0.145 & 0.982 \\ 

V4    & 16.715 & 0.034 & 45 & 0.177 & 0.415 \\ 
      &        &       &    & 0.038 & 0.218 \\ 
 
V6    & 16.696 & 0.123 & 24 & 0.079 & 0.194 \\ 
      &        &       &    & 0.034 & 0.904 \\ 
      &        &       &    & 0.091 & 0.214 \\ 
 
V7    & 16.622 & 0.028 & 47 & 0.205  & 0.718 \\ 
      &        &       &    & 0.0664 & 0.805 \\ 
      &        &       &    & 0.0344 & 0.964 \\ 
 
V8    & 16.773 & 0.043 & 46 & 0.205 & 0.050 \\ 
      &        &       &    & 0.045 & 0.603 \\ 
      &        &       &    & 0.031 & 0.905 \\ 
\hline 
\end{tabular} 
\end{table}

\begin{table} 
\caption[]{ Fourier decomposition of $I$-lightcurves} 
\label{Tab-fourier-I} 
\begin{tabular}{llllll} \hline 
 Name & $<I>$  & rms   & N  & Ampl   & phase  \\ \hline  
12197 & 15.379 & 0.021 & 66 & 0.140  & 0.370  \\ 
      &        &       &    & 0.050  & 0.209 \\ 
      &        &       &    & 0.0059 & 0.958 \\ 
 
12198 & 15.234 & 0.016 & 74 & 0.156  & 0.374  \\ 
      &        &       &    & 0.067  & 0.248 \\ 
      &        &       &    & 0.036  & 0.149 \\ 
      &        &       &    & 0.018  & 0.998 \\ 
 
12199 & 15.591 & 0.015 & 67 & 0.168  & 0.963  \\ 
      &        &       &    & 0.081  & 0.325 \\ 
      &        &       &    & 0.042  & 0.741 \\ 
      &        &       &    & 0.020  & 0.047 \\ 
      &        &       &    & 0.0085 & 0.515 \\ 
 
12200 & 15.593 & 0.104 & 15 & 0.231 & 0.768 \\ 
      &        &       &    & 0.168 & 0.945 \\ 
 
12202 & 15.383 & 0.050 & 50 & 0.140  & 0.653 \\ 
      &        &       &    & 0.067  & 0.832 \\ 
      &        &       &    & 0.055  & 0.823 \\ 
      &        &       &    & 0.020  & 0.921 \\ 
 
12203 & 15.441 & 0.025 & 67 & 0.146  & 0.931 \\ 
      &        &       &    & 0.063  & 0.341 \\ 
      &        &       &    & 0.021  & 0.773 \\ 
      &        &       &    & 0.012  & 0.145 \\ 
 
V7    & 15.295 & 0.013 & 18 & 0.125  & 0.645 \\ 
      &        &       &    & 0.055  & 0.591 \\ 
      &        &       &    & 0.028  & 0.886 \\ 
 
\hline 
\end{tabular} 
\end{table}

In addition to $B$, $V$ and $I$ data, single-epoch $JHK$ photometry
was collected from the \M\ {\it all-sky release} (Cutri et
al.~2003). Accurate coordinates are not immediately available in the
literature for the Cepheids, and therefore coordinates were retrieved
using a {\sc fits} image containing the necessary WCS (World
Coordinate System) keywords and the finding charts in AT67, Storm et
al. (1988) and We91. The photometry with errors and the coordinates,
as given by \M, are listed in Table~\ref{Tab-2mass}.  In general, the
stars are faint (for \M) and the error bars are substantial.
Monitoring these stars in the infrared with modern instrumentation
would be valuable. It is immediately evident that the photometry for
HV~12200 is very different from all others. Possibly, we have
identified the wrong star, or there is an unnoticed problem with the
\M\ photometry for this object. This star was not used when fitting
the $K$-band $PL$-relation.
 
\section{Analysis} 
 
\begin{figure} 
\centerline{\psfig{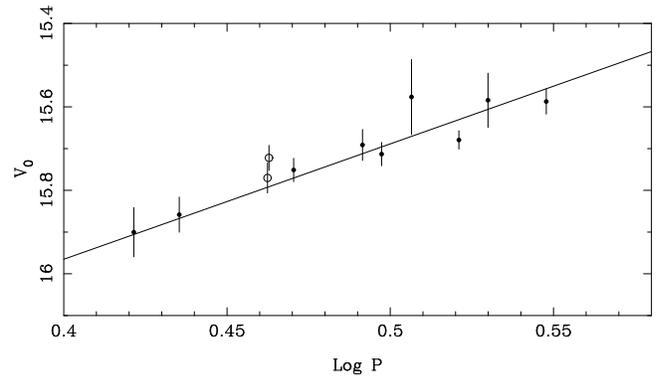}} 
\caption[]{ $PL$ relation in the $V$-band, for the adopted slope of 
$-2.765$ and $E(B-V)$ = 0.12 (see text for a discussion 
about the cluster reddening). The open circles represent the two overtone
pulsators, plotted at their corresponding fundamental period. }
\label{Fig-plv} 
\end{figure}

\begin{figure} 
\centerline{\psfig{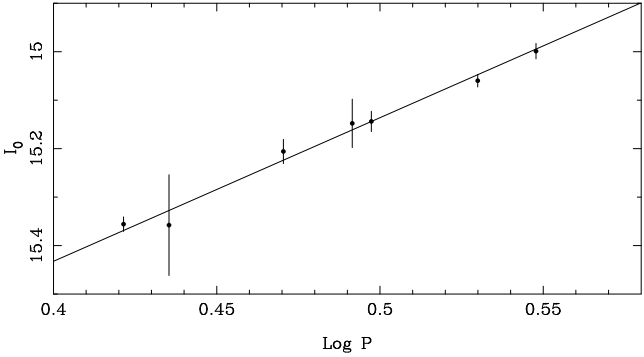}} 
\caption[]{ 
$PL$ relation in the $I$-band, for the adopted slope of $-2.963$ and $E(B-V)$ = 0.12. 
} 
\label{Fig-pli} 
\end{figure}

The previous study of the available Cepheid photometry has provided us
with mean $B$,$V$ magnitudes for 11 objects, $K$ single epoch
magnitudes for 10 objects, and $I$ mean magnitudes for 7 objects; from
the $V$ and $I$ magnitudes we computed the corresponding values of the
reddening independent Wesenheit index $W = I - 1.55 (V-I)$, like in
Udalski et al.~(1999a, hereafter U99). In Figs.~\ref{Fig-plv},
\ref{Fig-pli}, ~\ref{Fig-plb}, \ref{Fig-plw} we display the $PL$
relationships for the cluster Cepheids in $V$, $I$, $B$ and $W$; we
fitted to the data, as customary, a $PL$-relation of the type
\begin{equation} 
m_{\rm c} = ({\rm slope}) \times \log P + {\rm ZP} 
\end{equation} 
with the index $c$ being $W, I$, $B$ and $V$.  We obtained slopes
equal to $-3.47 \pm$ 0.48 in $W$, $-2.94 \pm$ 0.17 in $I$, $-2.40 \pm$
0.50 in $B$ and $-2.52 \pm$ 0.33 in $V$.  Due to the small number of
objects the error on the slopes is substantial, but it does still
allow interesting inferences.  First of all, we have compared these
slopes to the corresponding values for the LMC field Cepheids as
obtained by Groenewegen~(2000) from the data by U99, i.e., $-3.337$,
$-2.963$, $-2.352$\footnote{For this passband the slope was not given
in Groenewegen (2000) but has been determined for the present paper in an
identical way.} and $-2.765$ in $W$, $I$, $B$ and $V$ respectively,
with very small errorbars. It is evident that the cluster $PL$ slopes
are in formal agreement with the field ones.

We have then compared the cluster slopes with the recent results by
Tammann, Sandage \& Reindl~(2003) for Galactic Cepheids; the authors
re-calibrated the Galactic $PL$ relationship by combining absolute
magnitudes of Cepheids in open clusters (distances obtained from the
MS-fitting technique) with absolute magnitudes of other Cepheids
obtained from surface brightness methods, and obtained slopes equal to
$-2.757 \pm 0.112$, $-3.408 \pm 0.095$ and $-3.141 \pm 0.100$ in $B$,
$I$ and $V$, respectively.  These values are significantly different
from the results for the field LMC Cepheids, pointing out to a clear
dependence of the slopes of the $PL$ relationships on the Cepheid
metallicities.  It is also evident that the cluster $PL$ slopes are
significantly different from the Galactic ones, even accounting for
their associated large error.  Fouqu\'e, Storm \& Gieren~(2003)
provide an alternative calibration for the Galactic Cepheids slopes,
based on surface brightness methods: $-3.57 \pm$ 0.10 in $W$, $-3.24
\pm$ 0.11 in $I$, $-2.72 \pm$ 0.12 in $B$, and $-3.06 \pm$ 0.11 in
$V$; again, the slopes in $B$, $V$ and $I$ are different than in case of
the LMC field Cepheids, and are also different from the NGC~1866
results at more than 1$\sigma$ level.  The result that Galactic
Cepheids have a different $PL$ slopes compared to the LMC ones
contradicts the standard assumption of universality of the $PL$
relationship; if confirmed, this occurrence allows one to firmly
exclude a solar metallicity for NGC~1866 (one of the possible
explanations for the distance discrepancy mentioned in S03).

\begin{table*} 
\caption[]{ \M\ photometry} 
\label{Tab-2mass} 
\begin{tabular}{cccccc} \hline 
 Id   & RA (deg)  &  Dec (deg) &          $J$       &          $H$       & $K$              \\
\hline  
12197 & 78.342284 & -65.503525 & 14.776 $\pm$ 0.037 & 14.538 $\pm$ 0.068 & 14.423 $\pm$ 0.099\\ 
12198 & 78.361127 & -65.451439 & 14.816 $\pm$ 0.047 & 14.490 $\pm$ 0.061 & 14.469 $\pm$ 0.102\\ 
12199 & 78.354798 & -65.491463 & 14.958 $\pm$ 0.044 & 14.591 $\pm$ 0.080 & 14.407 $\pm$ 0.107\\ 
12200 & 78.440586 & -65.459991 & 14.322 $\pm$ 0.071 & 13.891 $\pm$ 0.079 & 13.696 $\pm$ 0.094\\ 
12202 & 78.452373 & -65.475998 & 14.638 $\pm$ 0.044 & 14.297 $\pm$ 0.073 & 14.153 $\pm$ 0.087\\ 
12203 & 78.459130 & -65.485962 & 14.913 $\pm$ 0.054 & 14.687 $\pm$ 0.084 & 14.335 $\pm$ 0.100\\ 
12205 & 78.576648 & -65.510002 & 14.587 $\pm$ 0.040 & 14.366 $\pm$ 0.056 & 14.254 $\pm$ 0.066\\ 
V4    & 78.406977 & -65.454140 & 14.746 $\pm$ 0.063 & 14.406 $\pm$ 0.076 & 14.520 $\pm$ 0.141\\ 
V6    & 78.424360 & -65.472099 & 15.019 $\pm$ 0.067 & 14.793 $\pm$ 0.108 & 14.650 $\pm$ 0.151\\ 
V7    & 78.426274 & -65.458389 & 14.810 $\pm$ 0.057 & 14.521 $\pm$ 0.081 & 14.391 $\pm$ 0.097\\ 
V8    & 78.428394 & -65.455498 & 14.981 $\pm$ 0.064 & 14.687 $\pm$ 0.080 & 14.637 $\pm$ 0.143 \\ 
\hline 
\end{tabular} 
\end{table*} 
 
As for the $PL$ relationship in $K$ (Fig.~\ref{Fig-plk}) the data are very 
much scattered, and do not allow a meaningful determination of the $PL$ 
slope in this photometric band. 
 
\begin{figure} 
\centerline{\psfig{figure=h4614f3.ps,width=8.5cm}} 
\caption[]{ $PL$ relation in the $B$-band, for the adopted slope of
$-2.352$ and $E(B-V)$ = 0.12. The open circles represent the two overtone
pulsators, plotted at their fundamental period. }
\label{Fig-plb} 
\end{figure} 
 
\begin{figure} 
\centerline{\psfig{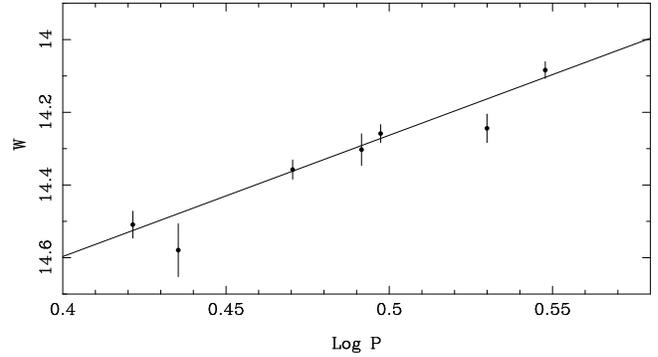}} 
\caption[]{ 
$PL$ relation for the Wesenheit index, for the adopted slope of $-3.337$. 
}
\label{Fig-plw} 
\end{figure} 
 
\begin{figure} 
\centerline{\psfig{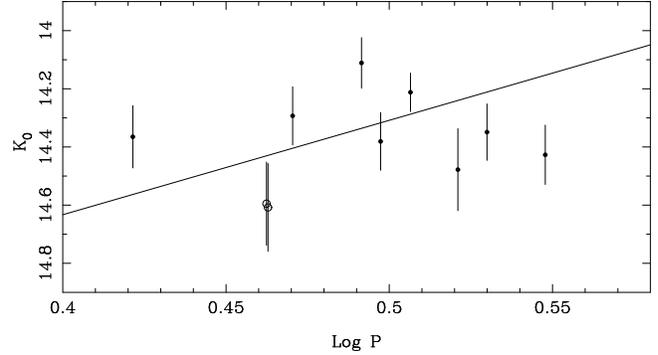}} 
\caption[]{ $PL$ relation in the $K$-band, for the adopted slope of 
$-3.246$ and $E(B-V)$ = 0.12.  The open circles represent the two overtone
pulsators, plotted at their fundamental period. }
\label{Fig-plk} 
\end{figure} 
 
The second step in our analysis has been the determination of the
distance of NGC~1866 from the LMC main body by using Cepheid stars.
Due to the statistical agreement between the $PL$ slopes (in $W$, $I$,
$B$ and $V$) for the cluster and the field, and the similarity between
the spectroscopic [Fe/H] determination for the cluster--[Fe/H] =
$-0.5 \pm 0.1$ according to Hill et al~(2000)--and the mean [Fe/H] of
LMC Cepheids and supergiants--[Fe/H]$\sim -$0.4 according to Luck \&
Lambert~(1992)--we can safely assume for the cluster Cepheids the very
accurate slopes obtained for the field LMC objects.  First, we fitted
the $W$ $PL$ relationship for the field LMC Cepheids to the cluster
objects; this relationship is reddening independent, therefore we do
not have to use any information regarding the cluster $E(B-V)$.
Adopting the ZP for the LMC field Cepheids from Groenewegen (2000), we
obtained a relative distance modulus of 0.04 $\pm$ 0.03 mag with
respect to the LMC one, implying that the cluster is slightly more
distant than the LMC main body.  In order to establish a connection
between this result and the distance between the cluster and its
surrounding LMC field, one needs a model for the geometry of the
LMC. There is a general consensus about the fact that the LMC is a
disk galaxy with an approximately planar geometry; the two basic
parameters to be evaluated are the inclination angle $i$ and the
position angle $\theta$ of the line of nodes (the intersection of the
galaxy plane and the sky plane), for which different estimates exist
in the literature, as reported in Table~\ref{geom}. These different
values for $i$ and $\theta$ imply different distances between the
field around NGC~1866 and the LMC centre, as displayed in
Table~\ref{geom}. For a distance modulus of $\sim$18.50 mag to the
galaxy centre, the distances reported in Table~\ref{geom} correspond to
a difference between $-$0.07 and $+$0.04 mag around this value.  By
interpreting the cluster distance modulus offset by 0.04$\pm$0.03 mag
with respect to the galaxy main body as the distance between NGC~1866
and the LMC centre, we obtain a difference of the distance moduli to
the cluster and the surrounding field $\Delta$(DM) ranging between
zero and $-$0.11 mag (the field being closer), depending on the
accepted values of $i$ and $\theta$.  This result definitely rules out
the possibility that the value $\Delta$(DM) = +0.20 $\pm$ 0.10 obtained
by S03 can be attributed to the cluster being closer.
 
\begin{table} 
\caption[]{The distance between the field around NGC~1866 and the LMC centre.} 
\label{geom} 
\begin{tabular}{cccccccc}  \hline 
$\theta$ & $i$     & Reference & $\Delta$ \\ 
 (\degr) & (\degr) &           & (kpc)    \\ 
\hline 
 258     & 38  & Schmidt-Kaler \& Gochermann (1992) &  0.37 \\ 
 258     & 33  & Feitzinger et al. (1977)           &  0.33 \\ 
 261     & 25  & Weinberg \& Nikolaev (2000)        &  0.38 \\ 
 296     & 18  & Groenewegen (2000)                 &  1.02 \\ 
 232     & 29  & Martin et al. (1979)               & -0.65 \\ 
 212     & 35  & van der Marel \& Cioni (2001)      & -1.61 \\ 
\hline 
\end{tabular} 
\end{table} 
 
At this point we can also redetermine the cluster reddening by 
imposing that the distances obtained from the $B$, $V$ and $I$ $PL$ 
relationships (which depend on the assumed cluster reddening) must 
provide the same relative distance from the LMC main body as obtained 
from $W$.  We do not use the $K$-band to derive $E(B-V)$, due to the 
large spread of the cluster data in this $PL$ plane and its weak 
sensitivity to the reddening; however, we will employ $K$ data (with 
the $PL$ slope fixed by the LMC field Cepheids, as determined by 
Groenewegen~2000) as a sanity check for the results obtained from $V$ 
and $I$. 
 
In our analysis we will use the following extinction ratios: 
 
$$A_B = 4.32 E(B-V)$$ 
$$A_V = 3.24 E(B-V)$$ 
$$A_I = 1.96 E(B-V)$$ 
$$A_K = 0.35 E(B-V)$$ 
 
\noindent 
following Schlegel et al.~(1998), homogeneously with the OGLE-II 
(Udalski et al.~1999b) extinction maps; we recall that the zero point 
of the LMC $PL$ relationships is derived from U99 data and the OGLE-II 
extinction maps, which provide an average reddening $E(B-V)$ = 0.15 for 
the LMC Cepheids. 

By assuming the canonical value $E(B-V)$ = 0.06, the cluster results
to be more distant than the LMC main body by 0.31 mag in $B$,
0.22 mag in $V$ and 0.18 mag in $I$. Obviously, this dependence on
colour suggests an incorrect reddening, and in fact we obtain
agreement between the distances in $B,V,I$ and $W$ only when the
reddening is higher than the canonical value. More in detail, we
compute various distances cluster--LMC main body, by fixing the
reddening each time at a different value. For each reddening selection
a distance with its associated 1$\sigma$ error is obtained. Typical
errors obtained from the fitting procedure are $\pm$0.02 mag in $I$,
$\pm$0.05 mag in $V$ and $\pm$0.07 mag in $B$; we then enforce the
condition that the 'true' cluster reddening has to provide the same
(within the respective error bars) relative distance in all of the
three photometric bands, equal to the value obtained from the
reddening free $W$ index. We obtain $E(B-V)$ = 0.12 $\pm$ 0.02, where
the error bar is basically determined by the more precise $I$-band data.

This is an important result, because it points out to a severe
underestimate of the cluster reddening, with relevant implications for
the MS-fitting distance.  This reddening is also consistent (in the
limit of the large dispersion of the $K$-band data) with the
constraint imposed by the $K$-band data.

\begin{figure} 
\centerline{\psfig{figure=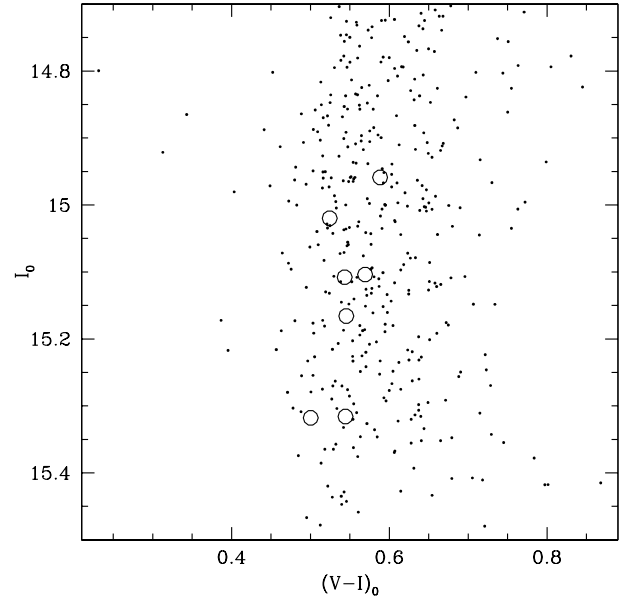,width=8.5cm}} 
\caption[]{ $I_0-(V-I)_0$ diagram for the cluster (open circles) and
FU LMC field (dots) Cepheids. A reddening $E(B-V)$ = 0.12 and a
correction of $-0.04$ mag in $I$ have been applied to the cluster data. 
OGLE-II reddenings have been employed for the field Cepheids.
}
\label{instrip} 
\end{figure}

We have also performed, as a consistency check, a comparison
between the positions of the cluster and field Cepheids in the
$(V-I)_0-I_0$ Colour Magnitude Diagram (the $V$- and $I$-band $PL$
relationships for the cluster Cepheids are the best defined ones), to
verify if the cluster Cepheids are within the Colour Magnitude Diagram
instability strip of the LMC field objects. Only FU field objects are
displayed, since all cluster Cepheids with $I$ photometry available
happen to be FU pulsators. We have used $E(B-V)$ = 0.12 for the
cluster, the OGLE-II reddenings for the field Cepheids, and we have
applied a correction of $-$0.04 mag to the $I$ magnitudes of the
cluster objects, to account for their distance to the main body of the
LMC derived before. Figure~\ref{instrip} displays the result of this
comparison, and shows clearly that $E(B-V)$ = 0.12 for NGC~1866 is
compatible with the location of the instability strip at the LMC
metallicity.

As a final step we can try to derive an absolute value for the
distance to NGC~1866 by applying a calibration for the absolute
magnitude of the zero point of the Cepheids $PL$ relationship. The
release of the $Hipparcos$ database has prompted a calibration of the
zero point based on parallaxes of Galactic Cepheids by means of the
reduced parallax method (see, e.g., Feast \& Catchpole~1997,
Groenewegen \& Oudmaijer~2000, Groenewegen~2000); the basic assumption
in this calibration is that the slope of the Galactic $PL$
relationships are the same as in the LMC, where they can be accurately
determined.  The results by Tammann et al.~(2003) and Fouqu\'e et
al.~(2003) mentioned before seem to clearly point out to a dependence
on metallicity, at least for the slope, whereas the situation about
the zero point is not clear; therefore LMC distance estimates obtained
with this kind of calibrations are affected by some uncertainty.
Nevertheless, we applied the Groenewegen~(2000) calibration of the
absolute magnitude zero point of the reddening independent Wesenheit
$PL$ relationship to our NGC~1866 sample, obtaining a cluster distance
modulus DM = 18.65 $\pm$ 0.10. Clearly, on the basis of the previous
discussion, the reddening dependent distances obtained from the $V$,
$I$ (and $K$) $PL$ relationships would provide a consistent result
when $E(B-V)$ = 0.12 $\pm$ 0.02 is assumed.  Based on the geometrical
corrections in Table~\ref{geom}, the field surrounding NGC~1866 is
located between DM = 18.65 and DM = 18.54. Notice how this last value
agrees well with the RC distance estimated by S03.

\section{Discussion} 
 
In the previous section we have obtained important results from the
analysis of NGC~1866 Cepheid population. First, we have been able to
derive a reddening independent relative distance between the cluster
and the surrounding field, that goes in the opposite way with respect
to what is necessary to explain the discrepancy found by S03.  Cluster
Cepheids provide a difference between the distance to the field around
NGC~1866 and the distance to the cluster, $\Delta$(DM), ranging
between zero and $-$0.11 (depending on the exact geometry of the LMC
disk), whereas S03 found $\Delta$(DM) = +0.20 $\pm$ 0.10.
 
Second, from the observed $PL$ relationships in $V$ and $I$ we have
obtained a new estimate of the cluster reddening, $E(B-V)$ =
0.12 $\pm$ 0.02, which is on the same scale as the OGLE-II extinction
maps of the LMC.  This reddening is about twice the canonical value
used for the cluster.
 
Third, in the assumption of universality of the Cepheid $PL$
relationships, we have obtained a cluster distance modulus DM =
18.65 $\pm$ 0.10.  Based on the geometrical corrections reported in
Table~\ref{geom}, this Cepheid distance implies that the field
surrounding NGC~1866 is located at a distance between DM = 18.65 and
DM = 18.54.
 
The new result about the cluster reddening has very important
implications for the distance discrepancy discussed in S03; first of
all, let us reexamine those results. The RC distance to the field
surrounding NGC~1866 derived by S03 is DM = 18.53 $\pm$ 0.07, and the
simultaneous reddening determination provided $E(B-V)$ = 0.05 $\pm$
0.02.  In a very recent paper Salaris, Percival \& Girardi~(2003b)
have studied in detail the systematic errors involved in the RC
distance method, when one takes into account current uncertainties in
the determination of the star formation history of the LMC, which is a
crucial input parameter for applying the method.  Based on Salaris et
al.~(2003b) results, one should revise the error bar on the previous
estimate, obtaining DM = 18.53 $\pm
0.07(random)^{+0.02}_{-0.05}(systematic)$ and $E(B-V) = 0.05 \pm
0.02(random)^{+0.06}_{-0.04}(systematic)$, where the systematic error
is due to the uncertainty in the LMC star formation history.  The
MS-fitting distance to the cluster is DM = 18.33 $\pm$ 0.08 when using
a reddening $E(B-V)$ = 0.064 $\pm$ 0.011 and the spectroscopic
metallicity [Fe/H] = $-0.5 \pm$ 0.1.
 
Our new determination of the cluster $E(B-V)$ provides a higher value,
which implies a longer cluster distance from the MS-fitting technique,
and therefore a potential solution to the distance problem found by
S03. In light of the importance of this issue, we have reexamined the
existing case for the canonical value $E(B-V)$ = 0.06 for the cluster.

\begin{figure} 
\centerline{\psfig{figure=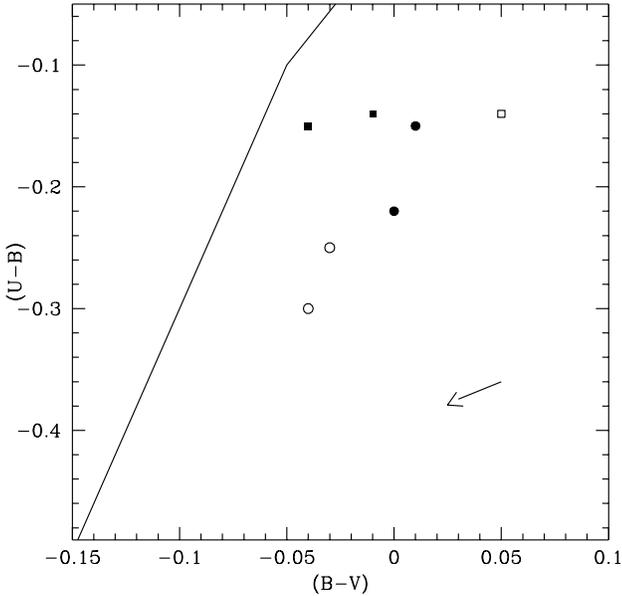,width=8.5cm}} 
\caption[]{ $(U-B)-(B-V)$ diagram with the photometry of the 4 objects 
used by Walker~(1974) to determine the cluster reddening.  Different 
symbols correspond to different stars; 3 out of 4 stars have 2 
observations.  The solid line represents the Pop~I standard Zero Age 
MS from Binney \& Merrifield~(1998).  The direction of the reddening 
vector is also displayed.  } 
\label{MScolcol} 
\end{figure}

The most direct empirical estimate of the cluster reddening before our
analysis was based on photoelectric $UBV$ photometry of 4 stars
observed by Walker~(1974).  The author compared the position of these
4 stars in the $(U-B)-(B-V)$ plane with a not clearly specified
standard Pop~I MS, and obtained (assuming $E(U-B)$ = 0.72$E(B-V)$) a
value quoted as $E(B-V)$ = 0.061 $\pm$ 0.0008 averaging over the
individual determinations made for the 4 stars (3 of them with 2
independent measurements, one with just one measurement); the formal
error is extremely small, and it is possibly due only to the
propagation of the internal error on the individual photometric
data. However, the individual reddening estimates show a dispersion of
$\sim$0.035 mag around this mean value, which we believe is a better
estimate of the error. Here, we have redetermined the cluster
reddening (see Fig.~\ref{MScolcol}) using the same method and cluster
data, but using the standard MS Pop~I sequence as reported in Table~3.9
of Binney \& Merrifield~(1998). We have preliminarily checked two
important effects. The first one is that the cluster Cepheids have
metallicity lower than solar, the second one is that the observed 4
stars are most likely evolved off the Zero Age MS, whereas the
standard Pop~I sequence represents the Zero Age MS. In order to check
the error introduced by these two factors, we used as a guideline the
behaviour of theoretical isochrones (Girardi et al.~2000) of ages
between 100 and 200 Myr (the typical age of the cluster) in the
$(U-B)-(B-V)$ plane, and found that both effects are practically
negligible for the objects observed in NGC~1866.  We then obtained
from the colour-colour diagram a reddening $E(B-V)$ = 0.08 $\pm$ 0.03,
comparable with the Cepheid estimate.  In addition, from the
$(B-V)$ and $(U-B)$ colours of one blue MS star reported by Wa87, we
obtained $E(B-V)$ = 0.10 $\pm$ 0.02 by applying the same method.

\begin{figure} 
\centerline{\psfig{figure=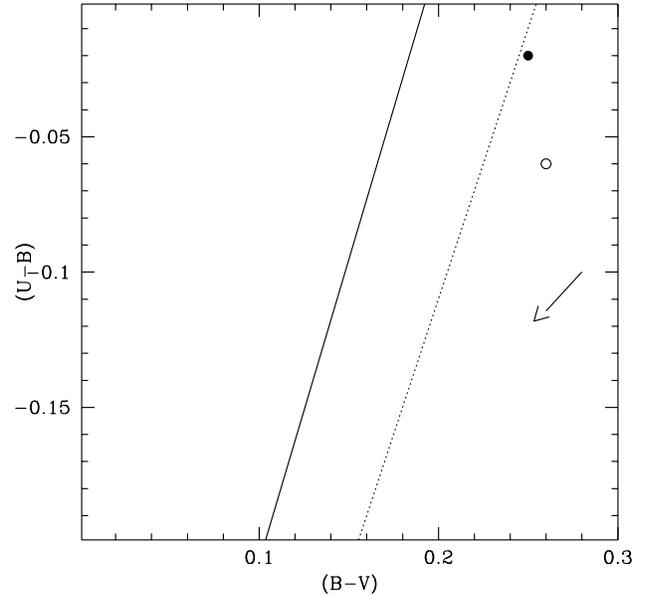,width=8.5cm}} 
\caption[]{ $(U-B)-(B-V)$ diagram for NGC~1866. The filled circle 
represents the colours by Bica et al.~(1996), the open circle the van 
den Bergh and Hagen~(1968) colours. The quoted photometric errors are 
equal to 0.02 mag in $(B-V)$ and $(U-B)$ for Bica et al.~(1996) data, 
0.02 mag in $(B-V)$ and 0.01 mag in $(U-B)$ for van den Bergh and 
Hagen~(1968) data. The solid line represents the Pop~I standard 
sequence by Schmidt-Kaler~(1967), the dotted line the standard 
sequence used by van den Bergh \& Hagen~(1968).  The direction of the 
reddening vector is also displayed.  } 
\label{clustcolcol} 
\end{figure}

The second evaluation used to support the canonical $E(B-V)$ = 0.06
was performed by van den Bergh \& Hagen~(1968). They determined the
reddening for a number of LMC clusters by comparing their integrated
colours with a standard sequence of intrinsic colours of Galactic open
clusters (based on results by Gray~1965 and Schmidt-Kaler~1967), in
the $(U-B)-(B-V)$ plane. They provide $E(B-V)$ = 0.06 for the cluster;
the error bar on the individual estimate is unspecified but they
clearly state that 'it should be emphasised that individual reddening
values are quite uncertain'. Moreover, they did not take into account
the effect of a possible metallicity difference between NGC~1866 and
Galactic open clusters.  We redetermined the cluster reddening using
this same procedure, and both the colours provided by van den Bergh \&
Hagen~(1968) and the recent redetermination by Bica et al.~(1996),
which is 0.01 mag bluer in $(B-V)$ and 0.04 mag redder in $(U-B)$.  We
have used two alternative colour-colour standard sequences for the
Galactic open clusters (see Fig.~\ref{clustcolcol}); the first one is
the relationship given by Eq.~(2) of Schmidt-Kaler~(1967), which
provides $E(B-V)$ = 0.14$\pm$0.02 and $E(B-V)$ = 0.10$\pm$0.02 from
the van den Bergh \& Hagen~(1968) and Bica et al.~(1996) colours,
respectively. In case of using the standard sequence reported in Eq.~(6)
of van den Bergh \& Hagen~(1968), we obtain $E(B-V)$ = 0.06$\pm$0.02
and $E(B-V)$ = 0.01$\pm$0.02. The error bar on the individual
determinations is due to the photometric error only, and not to the
error associated to the determination of the standard sequences.  It
has been possible also to estimate the effect of the chemical
composition, by using the theoretical results by Girardi et al.~(1995). 
In the hypothesis that the Galactic standard sequence has solar
metallicity, for a typical cluster age of 100 Myr and [Fe/H] = $-0.5$
the estimates given above should be reduced by about 0.02 mag.
 
It is evident, on the base of this discussion, that the cluster
integrated colours do not provide strong constraints on the cluster
reddening due especially to the uncertainty in the standard Galactic
sequence, and are not in clear contradiction with the reddening
obtained from the Cepheids.
 
As a further check we have compared the cluster reddening with
determinations of $E(B-V)$ for the surrounding field. As discussed
before, the multicolour RC method gives $E(B-V)$ =
0.05 $\pm 0.02(random)^{+0.06}_{-0.04}(systematic)$ which, within the
non negligible error bar, is not inconsistent with the Cepheid cluster
value.  We have in addition used the same OGLE-II technique to derive
the reddening for the field around the cluster. The method is based on
the assumption that the observed RC brightness in the $I$-band is
constant in the LMC (at least for the bar-inner disk fields observed
by OGLE-II) -- due to similar Star Formation Histories -- and
therefore differences in its apparent $I$ magnitude correspond to
extinction (hence reddening) differences when geometrical effects are
negligible or accounted for. The zero point of the reddening (with an
associated uncertainty by $\pm$0.02) is fixed by other independent
calibrators and confirmed by the recent work by Tammann et al.~(2003).
By comparing the RC apparent $I$ magnitude given by S03 for the field
around the cluster, with the OGLE-II fields close to the LMC centre,
we obtain $E(B-V)$ = 0.11 $\pm$ 0.02. An additional systematic error by
$^{+0.036}_{-0.020}$ has to be added to this value, due to the
uncertain geometric correction to the observed field, thus providing
$E(B-V)$ = 0.11$^{+0.04}_{-0.03}$, consistent with the cluster reddening.

All this analysis is clearly based on the assumption that the
extinction laws in the Galaxy and the LMC are the same in the $B$,
$V$, $I$ wavelength range. It is not completely clear if there are
differences for these photometric bands, but very recently Gordon et
al.~(2003) have published an average LMC extinction law which is
slightly different from the Galactic one used in our study. More in
detail, from Gordon et al.~(2003) paper one obtains $A_B = 4.41 E(B-V)$,
$A_V = 3.41 E(B-V)$ and $A_B = 2.14 E(B-V)$; with these ratios the Wesenheit
reddening free index becomes $W = I-1.69 (V-I)$ and one also finds that
$E(U-B) = 0.88 E(B-V)$.  We have therefore redetermined the cluster
reddening and the distance between the cluster and the LMC main body
by using this LMC extinction law. Of course, this test overestimates
the net effect of the LMC reddening law, because part of the
extinction towards the cluster and the LMC field is due to the Galaxy
(hence one should use the Galactic extinction law), and part to the
LMC internal extinction.

First of all, we have considered the Wesenheit index, and obtained a
cluster distance modulus relative to the LMC main body equal to
0.05 $\pm$ 0.03 mag, almost the same value as for the case of using
Galactic extinction ratios.  We have then re-estimated the cluster
reddening from its Cepheid population by using the same procedure
discussed in the previous section.  As a first step we redetermined
the zero points of the OGLE-II reddening maps using the Gordon et
al.~(2003) LMC extinction law; we found that the $E(B-V)$ zero points
are changed by at most 0.01 mag (in the direction of increased
reddening). We then corrected appropriately the individual field
Cepheid reddenings by considering a zero point E(B-V) higher by 0.01
mag and employing the LMC extinction ratios; we redetermined the field
$PL$ relationships, that show an unchanged slope and zero points
slightly brighter than what used in the previous chapter.  These
relationships have been then fitted to the cluster data 
(using again the above mentioned LMC reddening law) and, using the
same procedure as before, we obtained a cluster reddening that is
within 0.01 mag of the value 0.12$\pm$0.02 estimated before.

It appears therefore that changes in the extinction ratios consistent
with available observations do not influence appreciably our
determination of the reddening to the cluster. Also the estimates
based on the colour-colour diagrams discussed before are not altered
by more than 0.01 mag when using these LMC extinction ratios.
 
We can now conclude by studying the effect of this new reddening
determination on the MS-fitting distance modulus\footnote{Also
these results are not affected appreciably if we use the LMC average
extinction law by Gordon et al.~2003 instead of the Galactic one.}.
When the Cepheid-based $E(B-V)$ = 0.12 $\pm$ 0.02 is employed, and
assuming the same spectroscopic metallicity [Fe/H]$ = -0.5\pm 0.1$ as
in S03, the cluster DM is increased by 0.25 mag with respect to S03
results, giving DM = 18.58 $\pm$ 0.08; when compared to the RC
distance of the surrounding LMC field, this distance provides
$\Delta$(DM) = $-0.05 \pm 0.10(random)^{+0.02}_{-0.05}(systematic)$,
in agreement with the values inferred from the Cepheids plus the
geometrical corrections. The absolute values of the distance to the
LMC obtained from both the RC method and the cluster MS-fitting are
therefore consistent. It is interesting to notice that, within
the 1$\sigma$ error bar, this MS-fitting distance is consistent with
the result by Gieren et al.~(2000b), based on the infrared surface
brightness technique applied to one cluster Cepheid, which provides DM
= 18.42 $\pm$ 0.10. This latter determination, albeit based at the
moment on just one object, is largely insensitive to uncertainties in
the adopted reddening and extinction ratios, and therefore provides an
independent check for the consistency of our reddening and distance
estimates\footnote{Gieren et al.~(2000b) used a reddening of 
$E(B-V)$ = 0.07 in their analysis, and the $F_{\rm V} - (V-K)$ 
surface-brightness relation of Fouqu\'e \& Gieren 1997. When using our
derived reddening of $E(B-V)$ = 0.12 and the latest calibration by
Nordgren et al. 2002 we estimate the surface-brightness based DM 
will go up to 18.46.}. 

Our revised MS-fitting cluster distance is also in agreement, within
the corresponding 1$\sigma$ error bars, with the distance obtained
from the Wesenheit $PL$ relationship applied to the cluster Cepheid
population, in the assumption that it does not depend on the
metallicity. By fixing the $PL$ slope to the value observed in the LMC
and calibrating its zero point absolute magnitude on Galactic Cepheids
with $Hipparcos$ parallaxes (Groenewegen~2000), this $PL$ relationship
provides DM = 18.65 $\pm$ 0.10 for the cluster.  This implies that
within the period range spanned by the cluster Cepheids considered in
our analysis, the metallicity effects on the Wesenheit $PL$
relationship appear to be small, and probably in the direction of
slightly overestimating the cluster (which is on average more metal
poor than local Cepheids) distance.

\acknowledgements{ 
This research has made use of the SIMBAD database, operated at CDS, 
Strasbourg, France. 
This publication makes use of data products from the Two Micron All 
Sky Survey, which is a joint project of the University of 
Massachusetts and the Infrared Processing and Analysis 
Center/California Institute of Technology, funded by the National 
Aeronautics and Space Administration and the National Science 
Foundation. We thank an anonymous referee for his/her comments and
suggestions, which improved the presentation of our results.
}

{} 
 
\end{document}